\documentclass[reprint,superscriptaddress,amsmath,amssymb,aps,showpacs,prl]{revtex4-1}

\usepackage{bm}
\usepackage{amssymb}
\usepackage{amsfonts}
\usepackage{braket}
\usepackage{epsfig}
\usepackage{graphicx}
\usepackage{stmaryrd}
\usepackage{subfig}
\usepackage{natbib}
\usepackage{lipsum}

\newcommand{\vep}{\varepsilon}
\newcommand{\ua}{\uparrow}
\newcommand{\da}{\downarrow}
\newcommand{\la}{\langle}

\newcommand{\ra}{\rangle}

\newcommand{\sig}{\sigma}
\newcommand{\nn}{\nonumber}

\newcommand{\al}{\alpha}
\newcommand{\be}{\beta}
\newcommand{\ga}{\gamma}
\newcommand{\de}{\delta}

\newcommand{\om}{\omega}

\makeatletter
\newcommand*{\rom}[1]{\expandafter\@slowromancap\romannumeral #1@}
\makeatother
\graphicspath{{figures/}}

\begin{document}

\date{\today}
\title{Emergence of heavy quasiparticles from a massless Fermi sea:
  Optical conductivity}
\author{Hyunyong Lee}
\email[]{hyunyong.rhee@gmail.com}
\affiliation{Division of Advanced Materials Science, Pohang University
  of Science and Technology (POSTECH), Pohang 790-784, South Korea}

\author{S. Kettemann}
\email[]{s.kettemann@jacobs-university.de}
\affiliation{Division of Advanced Materials Science, Pohang University
  of Science and Technology (POSTECH), Pohang 790-784, South Korea}
\affiliation{School of Engineering and Science, Jacobs University
  Bremen, Bremen 28759, Germany} 

\begin{abstract}
  We study the density of states and the optical conductivity of a Kondo lattice
  which is immersed in a massless Dirac Fermi sea, as characterized by a linear
  dispersion relation. As a result of the hybridization $V$ with the
  $f$-electron levels, the pseudo-gap in the conduction band becomes duplicated
  and is shifted both into the upper and the lower quasiparticle
  band. We find that due to the linear dispersion of the Dirac
  fermions, the Kondo insulator gap is observable in the optical conductivity in 
  contrast to the Kondo lattice system in a conventional conduction
  band, and the resulting gap\,[$\Delta_{\rm gap}(T)$] depends on
  temperature. The reason is that the Kondo insulator gap is an {\it indirect
    gap} in conventional Kondo lattices, while it becomes a {\it direct gap} in
  the Dirac Fermi Sea. We find that the optical conductivity attains two peaks
  and is vanishing exactly at $2 b(T)V$ where $b$ 
  depends on temperature.
\end{abstract}

\maketitle


{\it Introduction.-} Recent developments in synthesis techniques have led to the
discovery of many transition metals and rare earth compounds, heavy fermion
materials, which have a quantum critical point separating a magnetically ordered
phase from a paramagnetic phase\,\cite{stewart1}. The exotic low temperature
physics of heavy fermion systems\,(intermetallics synthesized on the basis of
$f$-electron elements), where the conduction electrons act as particles with
``huge masses'' comparable to that of a proton\,\cite{stewart1, stewart2}, can
be understood in terms of rescaled quasiparticles which relate the strong
correlations to the properties of virtual single particles\,\cite{pines}. The
heavy masses directly affect the low temperature properties of materials such as
the electric resistivity from electron-electron scattering and the heat capacity.  

On the other hand, the low-energy excitation in graphene and topological
insulators, among others, are fermionic quasiparticles described by a
relativistic ``massless''\,Dirac fermion, as characterized by a linear
dispersion relation rather than usual non-relativistic Landau
quasiparticles\,\cite{novoselov, zhang, xia, hsieh1, hsieh2}. It is a
diametrically opposed example of the heavy mass fermion. An intriguing question
then arises, and deserves both theoretical and experimental studies : how does
the composite quasiparticle of the massless fermion and localized $f$-electron
behave? 

Actually, M. H{\"o}ppner\,{\it et al.} have recently studied the interplay
between Dirac fermions and heavy quasiparticles in the layered material ${\rm
  Eu} {\rm Rh}_2 {\rm Si}_2$ by means of the angle resolved
photoemission\,(ARPES)\,\cite{hoppner}. They observed a Dirac-like conical band
with an apex close to the Fermi level, which passes through an ${\rm Eu}$ $4f^6$
final-state multiplet. They reached the conclusion that massless and heavy
fermion quasiparticles may not only coexist but can also strongly interact in
such solids.   

The interband transition in the optical conductivity of the heavy fermion
systems has been intensively studied both experimentally\,\cite{garner,
  dordevic, mena} and theoretically\,\cite{coleman3, shim, weber}. The interband
transition takes place between one band with more $f$-electron character and
another with more conduction electron character, and thus it involves an energy
scale at least of the order of the Kondo temperature\,$T_K$ below which the
local moment is screened. In a Kondo lattice there is another scale, the Fermi
liquid coherence scale\,$T_c$ below which the composite heavy fermion
develops\,\cite{coleman2}. Since the peak position of the optical conductivity
could be related to $T_K$ and $T_c$, it is important to obtain detailed
information about low energy excitations near the Fermi level. Optical
conductivity studies have been a useful tool for this purpose and have provided
much information on Kondo insulators\,\cite{travaglini, kimura1994, bucher}.   

In this work, we study the density of states\,(DOS) and the optical conductivity
of the Kondo lattice system which is immersed in massless Dirac fermion
bath. The pseudo-gap in the conduction band becomes duplicated and is shifted
both into the upper and the lower quasiparticle band. We find that these
pseudogaps are always outside the Kondo insulator gap regardless of other
parameters such as $f$-electron level and bandwidth. Remarkably, we find that
due to the linear dispersion of the Dirac fermions, the Kondo insulator gap is
direct gap, and is therefore observable in the optical  conductivity. This is in
contrast to the Kondo lattice system in a conventional conduction band.

{\it Model.-} Let us begin with the Anderson lattice model which is given by 
\begin{eqnarray}
  H &=& \sum_{\bm{k},\sig} \vep_{\bm{k}} c^{\dag}_{\bm{k}\sig} c_{\bm{k}\sig} 
  + \vep_{d} \sum_{i,\sig} d^{\dag}_{i\sig} d_{i\sig} \nn\\
  &+& U \sum_{i,\sig} n^d_{i\ua} n^d_{i\da}
  + V \sum_{i,\sig} [\, c^{\dag}_{i\sig} d_{i\sig} + h.c. \,],
\end{eqnarray}
where $c_{\bm{k}\sig}$ is the annihilation operator of conduction electron with
momentum\,$\bm{k}$ and spin-index\,$\sig$, $c_{i\sig}$\,($d_{i\sig}$) is the
annihilation operator of conduction\,(localized $f$) electron at site\,$i$ and
spin-index\,$\sig$, $U$ is the on-site Coulomb repulsion, $V$ is the
hybridization energy between the conduction and $f$-electron band, and
$\vep_{\bm{k}}$ and $\vep_d$ denote the conduction band and the local energy
level of $f$-electron, respectively. In order to avoid confusion between the
$f$-level and Fermi level\,$\vep_F$, we use $\vep_d$, instead of $\vep_f$.  

Introducing a slave boson operator\,$b\,(b^{\dag})$ and a Lagrange
multiplier\,$\lambda_0$ to avoid a double occupancy in the limit $U
\rightarrow \infty$, and diagonalizing in momentum space, the
mean-field Hamiltonian takes the form\,\cite{coleman1, gellmann}, 
\begin{eqnarray}
  H_{{\rm MFT}} 
  = \sum_{\bm{k},\sig,\al=\pm} E_{\bm{k}}^{\al} 
  a_{\bm{k}\sig}^{\al\dag} a_{\bm{k}\sig}^{\al} - 2 N \lambda_0,
\end{eqnarray}
where $N$ is the number of sites in the lattice,
$
  d_{\bm{k}\sig}^{\dag} = \frac{1}{\sqrt{N}} \sum_{i} d_{i\sig} 
  e^{i\bm{k} \cdot \bm{r}_i},\nn
$
and $a_{\bm{k}\sig}^{\pm}$ are linear combinations of $c_{\bm{k}\sig}$ and
$d_{\bm{k}\sig}$, playing the role of quasiparticle operators corresponding the
momentum state eigenenergies  
$
  E_{\bm{k}}^{\pm} = \frac{1}{2} \Big[\, \vep_{\bm{k}} + \vep_d' \pm
  \sqrt{(\vep_{\bm{k}} - \vep_d')^2 + 4b^2 V^2} \,\Big],
 $
where $\pm$ denotes the upper and lower band, $b \equiv \la b^{(\dag)} \ra$ is
the order parameter of condensation of the slave bosons. It depends on
temperature and vanishes in the mean field approximation at $T_K$. $\vep_{d}'
= \vep_d + \lambda_0$ is the renormalized $f$-level. Notice that the quasiparticle
spectrum is gapped in the regime of $\om_{{\rm gap}}^- < \om < \om_{{\rm gap}}^+$,
where 
$
\om_{{\rm gap}}^{\pm} =
 \frac{1}{2} \Big[\, \vep_d' \mp D \pm \sqrt{(\vep_d' \pm D)^2 
   + 4b^2 V^2} \,\Big],
$
and $D$ is the half bandwidth of conduction electrons.

\begin{figure}[Ht!]
  \captionsetup[subfloat]{font = {bf,up}, position = top} 
  \subfloat[~~~~~~$\ga = 0$~~~~~~~~~~~~~~]
  {\includegraphics[width=0.25\textwidth]{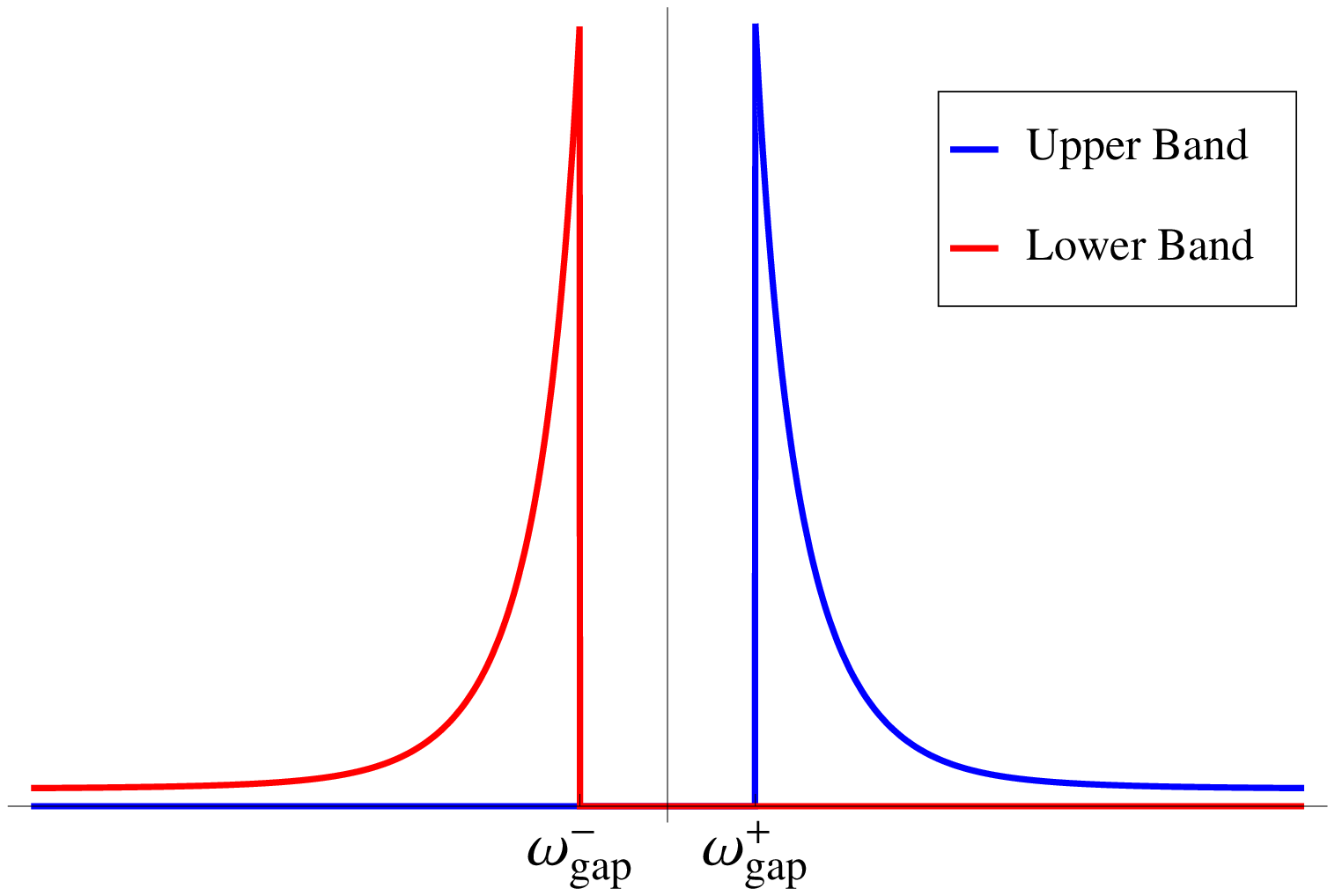}}
  \subfloat[~~~~~~$\ga = 1$~~~~~~~~~~~~~~]
  {\includegraphics[width=0.25\textwidth]{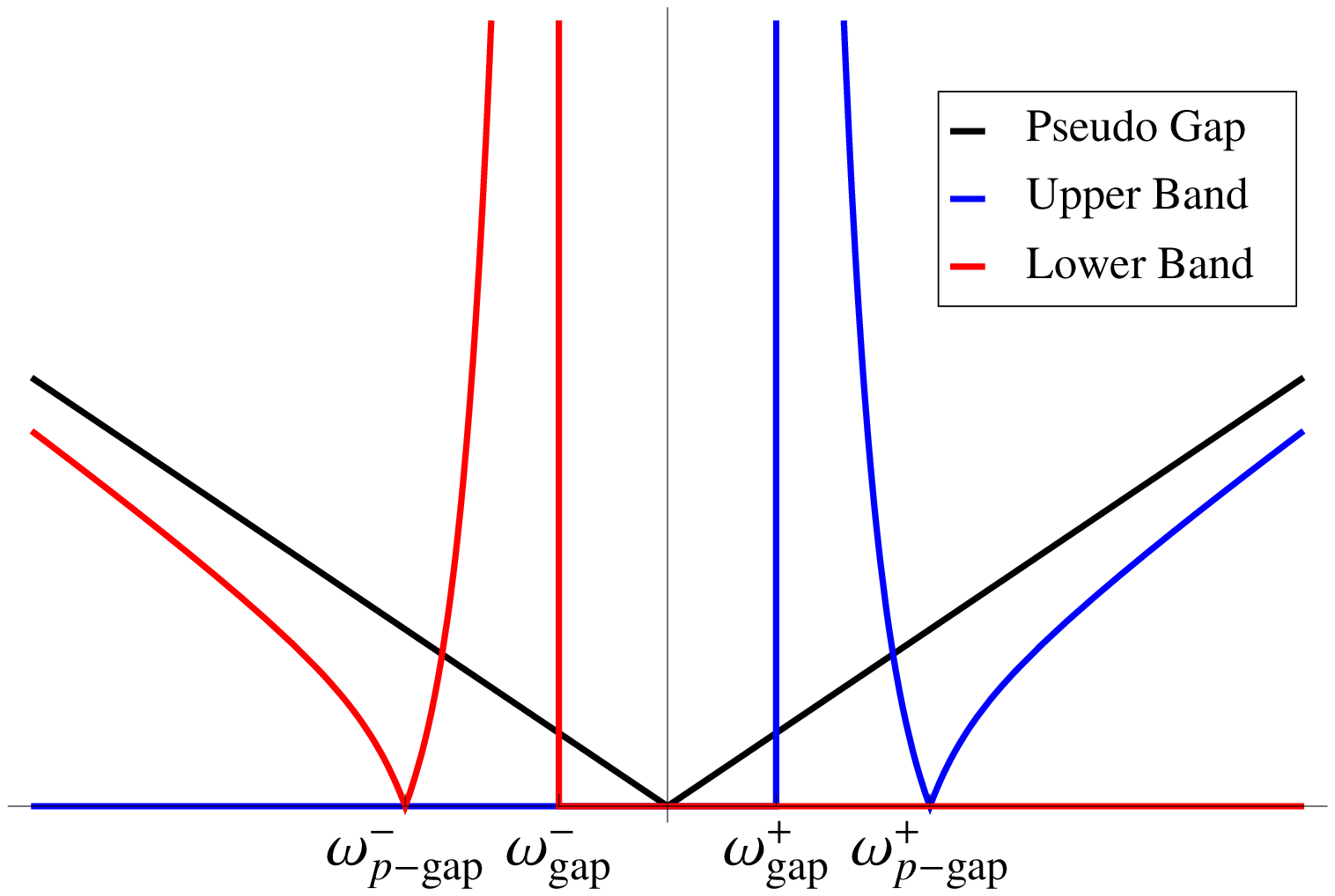}}
  \caption{(color online)  Quasiparticle density of states for
     pseudo-gap exponent (a) $\ga=0$\, and (b)
    $\ga=1$. Renormalized $f$-level and the position of pseudo-gap 
     in the conduction band are set to be
    at band center ($D = 1$ and $bV=0.1 D$). }    
  \label{fig:quasi_dos}
\end{figure}

{\it Density of states.-} One may calculate the density of states of the
quasiparticle from its dispersion, when the conduction band has a pseudo-gap,
$\rho_0(\om) = |\om|^{\ga}$ : 

\begin{eqnarray}
  \rho (\om) 
 = \sum_{\al=\pm} \frac{4\, \Theta(\al \,\om - \al \,
   \om_{{\rm gap}}^{\al} )}{1+ \al \frac{\om -\vep_d' - \frac{b^2 V^2}{\om 
        -\vep_d'}}{\sqrt{\Big( \om -\vep_d' 
        - \frac{b^2 V^2}{\om -\vep_d'} \Big)^2 +b^2 V^2}} } \, 
  \rho_0 (\om - \frac{b^2 V^2}{\om -\vep_d'}) \nn\\
  \label{eq:quasi_dos}
\end{eqnarray}
where $\Theta(\om)$ is the heavy side function. For $\ga=0$\,(flat band) and 
$\ga=1$\,(pseudo-gap), the quasiparticle DOS are plotted in
Fig.\,\ref{fig:quasi_dos}\,(a) and (b), respectively. As expected from previous
works\,\cite{auerbach, read1, read2}, the
originally flat conduction electron DOS\,($\ga=0$) is now replaced by a 
\textquotedblleft Kondo insulator gap\textquotedblright\,(also called
hybridization gap), flanked by two sharp peaks which are called coherence 
peaks\,[Fig.\,\ref{fig:quasi_dos}\,(a)]. 
Interestingly, the pseudo-gap in the conduction band is shifted to both upper and
lower bands after hybridization, so that two pseudo-gaps appear in the
quasiparticle DOS\,[Fig.\,\ref{fig:quasi_dos}\,(b)] at energies\,
$
  \om_{{\rm pseudo-gap}}^{\pm} = \frac{1}{2} \Big[\, \vep_d' \pm
  \sqrt{\vep_d'^2 + 4b^2 V^2} \,\Big],
$
which depend on the renormalized $f$-level and the effective hybridization $b
V$. These pseudogaps are found to be located always outside of the Kondo
insulator gap: $|\om_{{\rm pseudo-gap}}^{\pm}| > |\om_{{\rm gap}}^{\pm} |$. 

\begin{figure}[Ht!]
  \captionsetup[subfloat]{font = {bf,up}, position = top} 
  \subfloat[~~~~~~~~~$\vep_d = 0$~~~~~~~~~~~~~~~~~]
  {\includegraphics[width=0.25\textwidth]{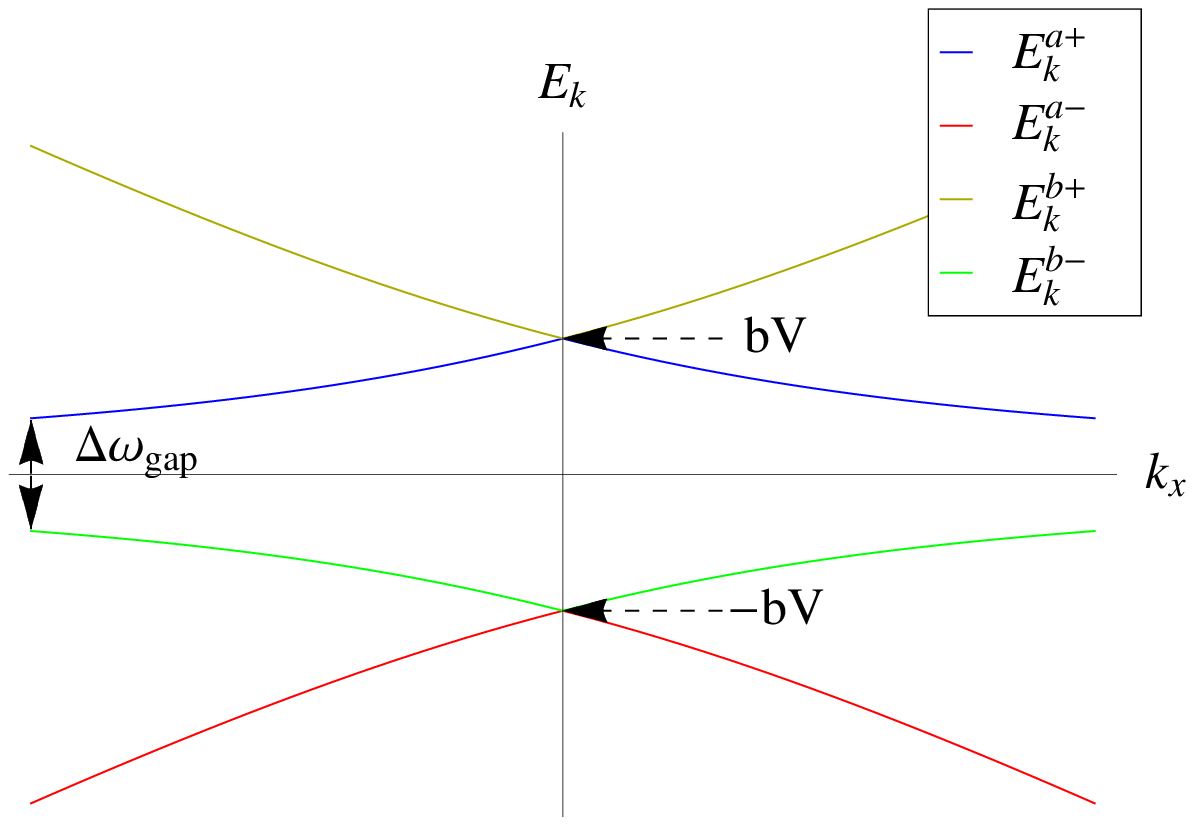}}
  \subfloat[~~~~~~$\vep_d = -0.4D$~~~~~~~~~~~~~~]
  {\includegraphics[width=0.25\textwidth]{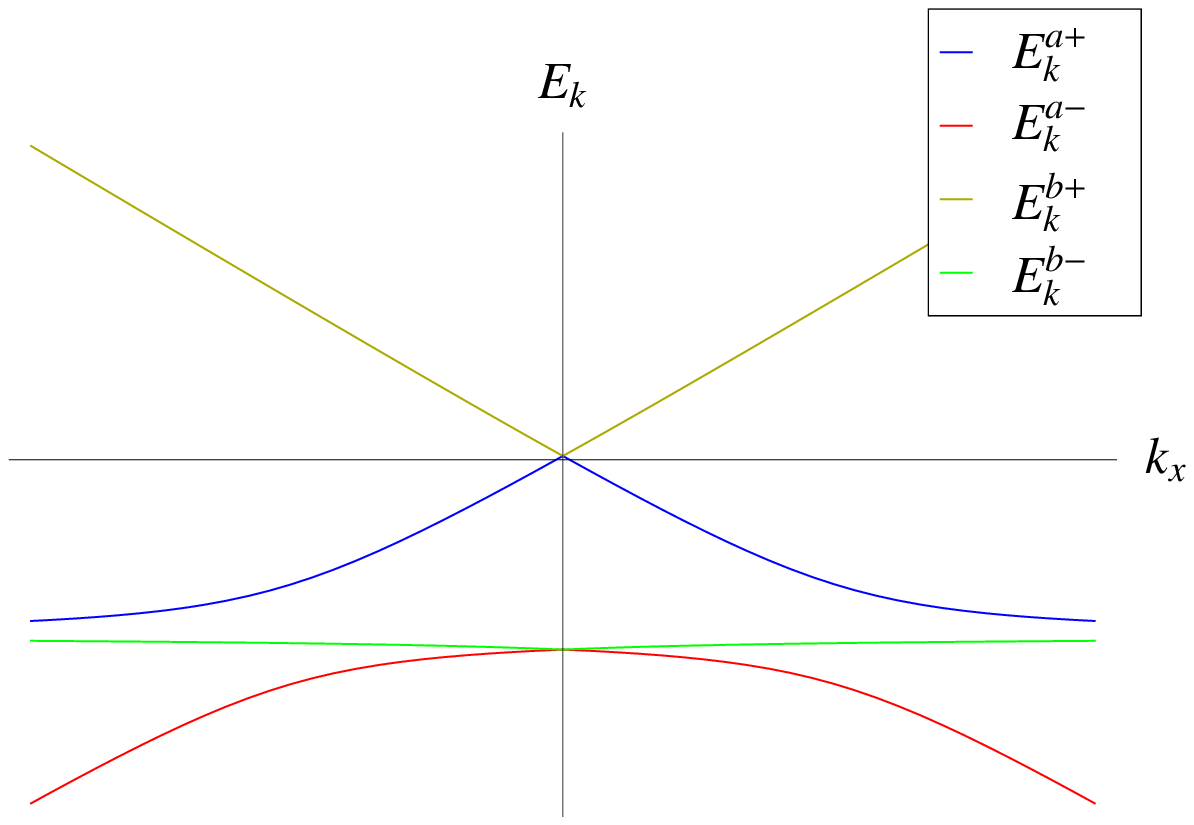}}
  \caption{(color online) 
   Quasiparticle   dispersion for (a) $\vep_d = 0$ and (b) $\vep_d = -0.4D$. 
    $bV=0.2 D$, $D$ and $v_F$ are set to unity.}    
  \label{fig:dirac_kondo}
\end{figure}

It is very well known that the electrons on a honeycomb lattice can be described
by a relativistic massless Dirac fermion model\,\cite{semenoff}, which is
characterized by the linear low energy dispersion relation $\vep_{\bm{k}} = \pm
v_F |\bm{k}|$ and $v_F$ is the Fermi velocity. S. Saremi and Patrick A. Lee have
derived the dispersion of the quasiparticles of 
the Kondo lattice model on a honeycomb
lattice, finding \,\cite{saremi} 
\begin{eqnarray}
  E_{\bm{k}}^{\alpha \pm} &=& \frac{1}{2} \Big\{ - \alpha v_F |\bm{k}| + \vep_d \pm 
  \sqrt{ (\alpha v_F|\bm{k}| + \vep_d)^2 + 4b^2 V^2 } \Big\}, \nn
  \label{eq:quasi_disp_honeycomb}
\end{eqnarray}
where we denote  $E_{\bm{k}}^{a \pm}$ for $\alpha =+$ and $E_{\bm{k}}^{b \pm}$
for $\alpha =-$ in Fig.\,\ref{fig:dirac_kondo}. When $\vep_d = 0$, the
pseudopgaps are at  two points outside the hybridization
gaps\,[Fig.\,\ref{fig:dirac_kondo}\,(a)], consistent with the DOS,
Fig.\,\ref{fig:quasi_dos}\,(b). When $\vep_d = -0.4D$, a Dirac-like cone is
preserved at the Fermi level, while a very  heavy quasiparticle is formed around
the $f$-level\,[Fig.\,\ref{fig:dirac_kondo}\,(b)]. Interestingly, similar
spectra have been reported in recent ARPES measurement of ${\rm EuRh}_2{\rm
  Si}_2$ known as an isostructural compound of ${\rm YbRh}_2{\rm
  Si}_2$\,\cite{hoppner}.

{\it Optical conductivity.-} The optical conductivity in the isotropic
system\,\cite{mahan} is defined by 
\begin{eqnarray}
  \sig(q,i\om) &=& \frac{1}{\om} {\rm Im} [\Pi (q, \om)], \\
  \Pi(\bm{q}, \om) &=& -\frac{1}{d} \int_0^{\be} d\tau e^{i \om \tau} 
  \la T_{\tau} \bm{j}(\bm{q}, \tau) \cdot \bm{j} (-\bm{q}, 0) \ra ,
\end{eqnarray}
where $\bm{j}(\bm{q}) = \frac{e}{m} \sum_{\bm{k}, \sig} (\bm{k} + \frac{1}{2} \bm{q})
c_{\bm{k}+\bm{q} \sig}^{\dag} c_{\bm{k} \sig}$ is the current operator, $d$ the
dimension, $e$, $m$ are the electron charge and  mass. $\Pi(\bm{q}, \om)$ is the
current-current correlation function, which in the bubble approximation and long
wavelength limit\,($\bm{q} \rightarrow 0$) is given by   
\begin{eqnarray}
  &&\Pi(i\nu_n)
  = \frac{G_0}{\Omega}  
  \frac{1}{\be} \sum_{\bm{k},i\om_n} v_F^2 G_c(\bm{k}, i\om_n) 
  G_c ( \bm{k}, i\om_n + i\nu_n ), 
  \label{eq:curcur_func}
\end{eqnarray} 
where  $G_0 = 2e^2/h$ is the conductance quantum, $\Omega$ is a volume of the
system, $\om_n$\,($\nu_n$) are fermionic\,(bosonic) Matsubara frequencies, and
$G_c(\bm{k},i\om_n)$ is the Green's function of conduction electrons in the
Kondo lattice system. 

We present the optical conductivity of a Kondo lattice in a conventional
uniform conduction band, in Sec.\,\rom{1} in Suppl., comparing with
Ref.\,\cite{paul,weber}. The threshold frequency\,$2bV$  is always larger than
the  Kondo insulator gap \,, $2bV > \om_{{\rm gap}}^+ - \om_{{\rm gap}}^- =
\Delta_{\rm gap}$, as can be understood  from the band structure of the
quasiparticles with an indirect band gap\,\cite{garner, dordevic, mena,
  coleman3, shim, weber}. Optical absorption is only possible for direct
transitions of the quasiparticle from lower band\,($E_{\bm{k}}^-$) to the  upper
band  with same momentum\,($E_{\bm{k}}^+$) due to  momentum conservation. Thus,
the smallest optical excitation\,(threshold) energy $2bV$ is larger than the
Kondo insulator gap\,$\Delta_{\rm gap}$ in a conventional Kondo lattice system.
%
%

On the other hand, from the dispersion of  the quasiparticles in a Dirac fermion
conduction bath\,[Fig.\,\ref{fig:dirac_kondo}], we may infer that the Kondo
insulator gap is observable in  the optical conductivity. The conduction
electron Green's function in  momentum and Matsubara frequency representation is
\begin{eqnarray}
  G_c(\bm{k}, i\om_n)
  &=& \sum_{\eta=\pm} \Bigg\{
  \frac{\al_{\bm{k}}^{a \eta}}{i\om_n - E_{\bm{k}}^{a \eta}}
  + \frac{\al_{\bm{k}}^{b \eta}}{i\om_n - E_{\bm{k}}^{b \eta}} \Bigg\},
  \label{eq:green_dirac}
\end{eqnarray}
where the quasiparticle eigenenergies\,$E_{\bm{k}}^{\al\eta}$ are defined in
Eq.\,\eqref{eq:quasi_disp_honeycomb} and the coherence factor is
$ 
\al_{\bm{k}}^{\al\pm} =\frac{E_{\bm{k}}^{\al\pm} - \vep_d'} {E_{\bm{k}}^{\al\pm}
  - E_{\bm{k}}^{\al\mp}}. 
$

Performing the Matsubara frequency summation in the current-current
correlation function \,Eq.\,\eqref{eq:curcur_func} for $T=0$, $\bm{q}
\rightarrow 0$ and $\om > 0$\,(see section\,\rom{2} in Suppl.),
we obtain 
 \begin{eqnarray}
   \sig(\om) = \frac{G_0}{\om} \sum_{\bm{k},\al, \be}
   v_F^2 \al_{\bm{k}}^{\al -} \al_{\bm{k}}^{\be +} \,\de( \om - E_{\bm{k}}^{\al +} +
   E_{\bm{k}}^{\be -}) ,
  \label{eq:opt_cond_dirac}
\end{eqnarray}
where $\al,\,\be = a\,{\rm or}\,b$. Note that there are 4 different kinds of
interband transitions in Eq.\,\eqref{eq:opt_cond_dirac}, and it may be expected
that the transition between the band $E_{\bm{k}}^{b-}$ and
$E_{\bm{k}}^{a+}$[blue and green line in Fig.\,\ref{fig:dirac_kondo}\,(a)]
allows us to observe the Kondo insulator gap of the quasiparticle DOS. After the
momentum summation, we acquire
\begin{widetext}
  \begin{eqnarray}
  \sig(\om) &=& \frac{\sig_0 \Theta(\om - 2bV)}{\om} \Bigg\{
  \Theta\Big(\frac{\Delta_{{\rm gap}}^2 + 4b^2 V^2}{2\Delta_{{\rm gap}}} - \om \Big) 
  \frac{\rho(\om_-) }{\Big| 1 - \frac{b^2 V^2}{\om_-^2} \Big|} 
  \frac{b^2 V^2 \om_-^2}{(\om_-^2 + b^2 V^2)^2}  
  +\Theta\Big(\frac{D^2 + 4b^2 V^2}{2D} - \om\Big) 
  \frac{\rho(\om_+) }{\Big| 1 - \frac{b^2 V^2}{\om_+^2} \Big|} 
  \frac{b^2 V^2 \om_+^2}{(\om_+^2 + b^2 V^2)^2}\nn\\
  &&~~~~~~~~~~~~~~~~~~~~
  +\Theta(D - \om) \frac{(\om/2)^4 \rho(\om/2)}{2\{(\om/2)^2 + b^2 V^2 \}^2} \Bigg\}
  +\frac{\sig_0 \Theta(\om - \Delta_{{\rm gap}}) \Theta(2bV - \om)}{2 \om}
  \frac{(\om/2)^4 \rho(\om/2)}{\{(\om/2)^2 + b^2 V^2 \}^2},
  \label{eq:opt_cond_dirac_final}
\end{eqnarray}
\end{widetext}
where $\rho(\om)$ is the quasiparticle DOS\,[Eq.\,\eqref{eq:quasi_dos} with
$\ga=1$], $\om_{\pm} = \frac{1}{2} \big\{ \om \pm \sqrt{\om - 4b^2 V^2} \big\}$, 
and $D$ is the half bandwidth. 
\begin{figure}[Ht!]
  \captionsetup[subfloat]{font = {bf,up}, position = top} 
  \subfloat[~~~~~~~~~~~~~~~~~~$\vep_d = 0$~~~~~~~~~~~~~~~~~~~~]
  {\includegraphics[width=0.35\textwidth]{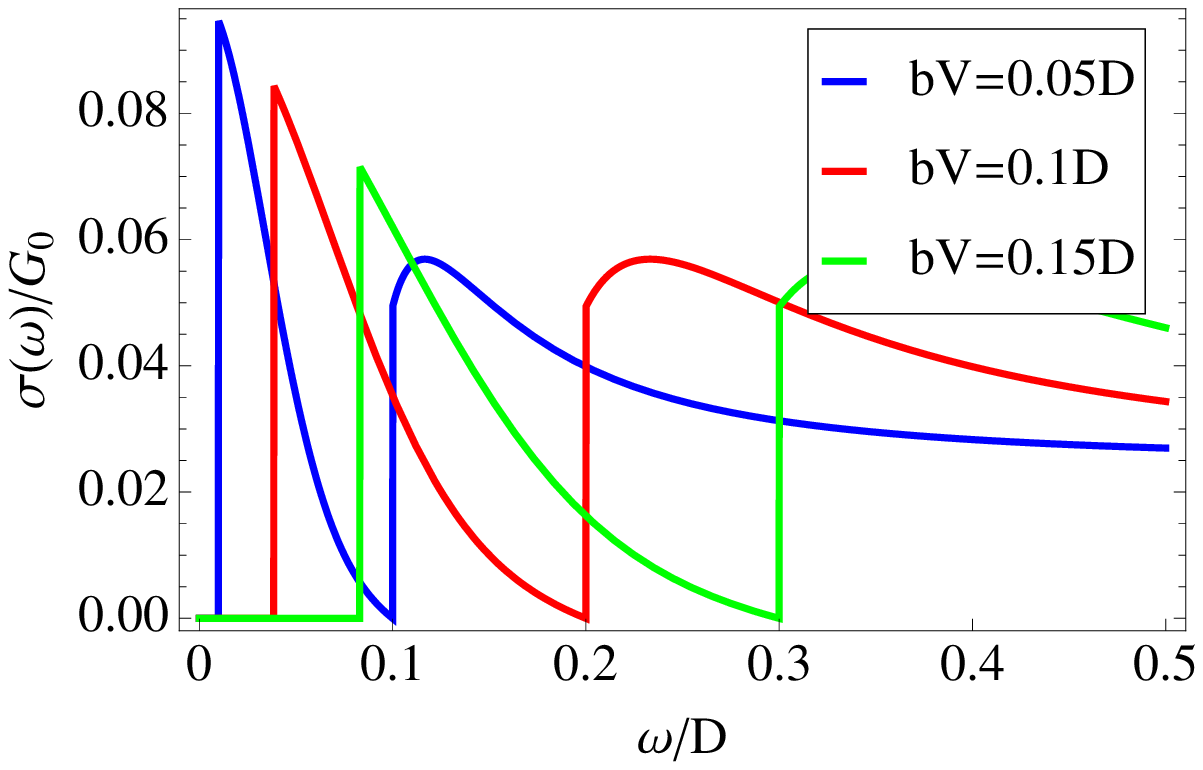}}\\
  \subfloat[~~~~~~~~~~~~~~~~$\vep_d = -0.4D$~~~~~~~~~~~~~~~~~]
  {\includegraphics[width=0.35\textwidth]{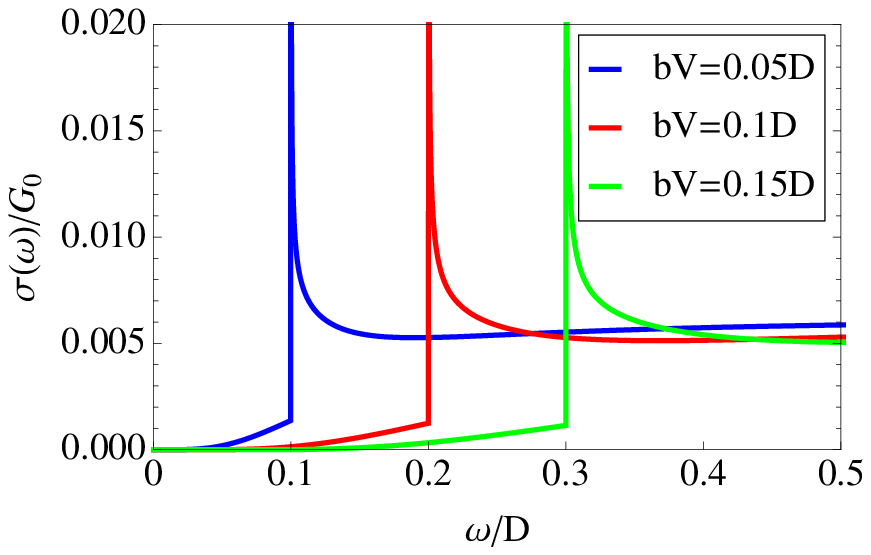}}
  \caption{(color online) Optical conductivity of Kondo lattice immersed in
    Dirac fermion\,($\ga=1$) for (a) $\vep_d = 0$ and (b) $\vep_d =
    -0.4D$. $v_F$ and $D$ are set to unity. The first peak appears at $\om =
    \Delta_{\rm gap}$, and it vanishes as the frequency approaches $\om = 2bV$ 
    where the second peak emerges.}  
  \label{fig:opt_cond_dirac}
\end{figure}
Fig.\,\ref{fig:opt_cond_dirac} shows the optical conductivity\,
Eq.\,\eqref{eq:opt_cond_dirac_final} for $\vep_d = 0,\,-0.4D$ for three different values
of $bV$. The Kondo insulator gap\,$\Delta_{{\rm gap}}$ is observable. The
optical conductivity is found to have 2 peaks contrary to the one in the
uniform\,($\ga=0$) conduction band system. While the first peak at the Kondo
insulator gap comes from the interband transition between the heavy $f$-electron
character quasiparticles, the second peak at $\om=2bV$ originates from the
transition between the Dirac-like quasiparticles around the pseudo-gaps. Notice
that due to the temperature dependence of $b=b(T)$, the Kondo insulator gap
depends on temperature: $\Delta_{\rm gap} = \Delta_{\rm gap}(T)$. We find that
the optical conductivity vanishes as the frequency is approaching
$\om\rightarrow 2bV$, since there is no state which can be excited from the
lower band to upper band as one can see from the energy
dispersion\,[Fig.\,\ref{fig:dirac_kondo}\,(a)]. When $\om > 2bV$, however, there
are transitions  from the lower to the  upper band so that the optical
conductivity has a finite value as a second peak. On the other hand, when the $\vep_d$
is far below the Dirac point\,[Fig.\,\ref{fig:opt_cond_dirac}\,(b)], the optical
conductivity changes drastically that it does not show a peak at $\Delta_{\rm
  gap}$, but increases as approaching $2bV$. Even though only a single 
peak appears like the one in the conventional\,($\ga=0$) conduction band system,
the threshold frequency is still $\Delta_{\rm gap}$ and the Kondo insulator gap
is observable as in the case of $\vep_d = 0$.

{\it Conclusions.-} In summary, we have derived the density of states and the
optical conductivity of a Kondo lattice immersed in a relativistic massless
Dirac fermion conduction bath, characterized by a linear dispersion. The
hybridization moves the pseudo-gap both in the upper and the lower quasiparticle
band. Due to the linear dispersion of the Dirac fermions, the Kondo insulator
gap is observable in the direct interband transition yielding the additional
peak in the optical conductivity at $\om = \Delta_{{\rm gap}}(T) \ll 2bV$. This
is in contrast, to the conventional  Kondo lattice  which   has a single peak at
$\om = 2bV$. While the first peak at $\om=\Delta_{\rm gap}$ comes from the
interband transition between the dispersionless $f$-electron character carriers,
the second peak at $\om=2bV$ originates from the Dirac-like massless conduction
electron character carriers. 

We would like to suggest experimental studies on the optical conductivity of the
heavy fermion compounds like ${\rm Eu} {\rm Rh}_2 {\rm Si}_2$\,\cite{hoppner}
showing the interference between the massless quasiparticles and localized
$f$-electrons. Also, magnetic layered topological insulators where ${\rm Co}$ is
deposited on a ${\rm Bi}_2{\rm Se}_3$ surface, and where the magnetic
properties reveal an absence of long range magnetism\,\cite{kimura2012}, might
be a good material to measure its optical conductivity and compare it with our
results. We believe that such experimental works may provide the first direct
observation of the Kondo insulator gap in the optical conductivity. Moreover,
such an experiment may allow to extract much information on the Kondo lattice
systems such as the Kondo temperature and the coherence temperature.  

\acknowledgments
We thank Ki-Seok Kim for helpful  and stimulating discussions. This
study was supported by the BK21 Plus funded by the Ministry of
Education, Korea (10Z20130000023).

\bibliographystyle{apsrev}
\bibliography{reference}

\begin{thebibliography}{29}
\expandafter\ifx\csname natexlab\endcsname\relax\def\natexlab#1{#1}\fi
\expandafter\ifx\csname bibnamefont\endcsname\relax
  \def\bibnamefont#1{#1}\fi
\expandafter\ifx\csname bibfnamefont\endcsname\relax
  \def\bibfnamefont#1{#1}\fi
\expandafter\ifx\csname citenamefont\endcsname\relax
  \def\citenamefont#1{#1}\fi
\expandafter\ifx\csname url\endcsname\relax
  \def\url#1{\texttt{#1}}\fi
\expandafter\ifx\csname urlprefix\endcsname\relax\def\urlprefix{URL }\fi
\providecommand{\bibinfo}[2]{#2}
\providecommand{\eprint}[2][]{\url{#2}}

\bibitem[{\citenamefont{Stewart}(2001)}]{stewart1}
\bibinfo{author}{\bibfnamefont{G.~R.} \bibnamefont{Stewart}},
  \bibinfo{journal}{Rev. Mod. Phys.} \textbf{\bibinfo{volume}{73}},
  \bibinfo{pages}{797} (\bibinfo{year}{2001}).

\bibitem[{\citenamefont{Stewart}(1984)}]{stewart2}
\bibinfo{author}{\bibfnamefont{G.~R.} \bibnamefont{Stewart}},
  \bibinfo{journal}{Rev. Mod. Phys.} \textbf{\bibinfo{volume}{56}},
  \bibinfo{pages}{755} (\bibinfo{year}{1984}).

\bibitem[{\citenamefont{Pines}(1999)}]{pines}
\bibinfo{author}{\bibfnamefont{D.}~\bibnamefont{Pines}},
  \emph{\bibinfo{title}{Elementary Excitations in Solids: Lectures on Protons,
  Electrons, and Plasmons}}, Advanced Book Classics
  (\bibinfo{publisher}{Advanced Book Program, Perseus Books},
  \bibinfo{year}{1999}), ISBN \bibinfo{isbn}{9780738201153}.

\bibitem[{\citenamefont{Novoselov et~al.}(2005)\citenamefont{Novoselov, Geim,
  Morozov, Jiang, Grigorieva, Dubonos, and Firsov}}]{novoselov}
\bibinfo{author}{\bibfnamefont{K.}~\bibnamefont{Novoselov}},
  \bibinfo{author}{\bibfnamefont{A.~K.} \bibnamefont{Geim}},
  \bibinfo{author}{\bibfnamefont{S.}~\bibnamefont{Morozov}},
  \bibinfo{author}{\bibfnamefont{D.}~\bibnamefont{Jiang}},
  \bibinfo{author}{\bibfnamefont{M.~K.~I.} \bibnamefont{Grigorieva}},
  \bibinfo{author}{\bibfnamefont{S.}~\bibnamefont{Dubonos}}, \bibnamefont{and}
  \bibinfo{author}{\bibfnamefont{A.}~\bibnamefont{Firsov}},
  \bibinfo{journal}{nature} \textbf{\bibinfo{volume}{438}},
  \bibinfo{pages}{197} (\bibinfo{year}{2005}).

\bibitem[{\citenamefont{Zhang et~al.}(2009)\citenamefont{Zhang, Liu, Qi, Dai,
  Fang, and Zhang}}]{zhang}
\bibinfo{author}{\bibfnamefont{H.}~\bibnamefont{Zhang}},
  \bibinfo{author}{\bibfnamefont{C.-X.} \bibnamefont{Liu}},
  \bibinfo{author}{\bibfnamefont{X.-L.} \bibnamefont{Qi}},
  \bibinfo{author}{\bibfnamefont{X.}~\bibnamefont{Dai}},
  \bibinfo{author}{\bibfnamefont{Z.}~\bibnamefont{Fang}}, \bibnamefont{and}
  \bibinfo{author}{\bibfnamefont{S.-C.} \bibnamefont{Zhang}},
  \bibinfo{journal}{Nature Physics} \textbf{\bibinfo{volume}{5}},
  \bibinfo{pages}{438} (\bibinfo{year}{2009}).

\bibitem[{\citenamefont{Xia et~al.}(2009)\citenamefont{Xia, Qian, Hsieh, Wray,
  Pal, Lin, Bansil, Grauer, Hor, Cava et~al.}}]{xia}
\bibinfo{author}{\bibfnamefont{Y.}~\bibnamefont{Xia}},
  \bibinfo{author}{\bibfnamefont{D.}~\bibnamefont{Qian}},
  \bibinfo{author}{\bibfnamefont{D.}~\bibnamefont{Hsieh}},
  \bibinfo{author}{\bibfnamefont{L.}~\bibnamefont{Wray}},
  \bibinfo{author}{\bibfnamefont{A.}~\bibnamefont{Pal}},
  \bibinfo{author}{\bibfnamefont{H.}~\bibnamefont{Lin}},
  \bibinfo{author}{\bibfnamefont{A.}~\bibnamefont{Bansil}},
  \bibinfo{author}{\bibfnamefont{D.}~\bibnamefont{Grauer}},
  \bibinfo{author}{\bibfnamefont{Y.}~\bibnamefont{Hor}},
  \bibinfo{author}{\bibfnamefont{R.}~\bibnamefont{Cava}}, \bibnamefont{et~al.},
  \bibinfo{journal}{Nature Physics} \textbf{\bibinfo{volume}{5}},
  \bibinfo{pages}{398} (\bibinfo{year}{2009}).

\bibitem[{\citenamefont{Hsieh et~al.}(2008)\citenamefont{Hsieh, Qian, Wray,
  Xia, Hor, Cava, and Hasan}}]{hsieh1}
\bibinfo{author}{\bibfnamefont{D.}~\bibnamefont{Hsieh}},
  \bibinfo{author}{\bibfnamefont{D.}~\bibnamefont{Qian}},
  \bibinfo{author}{\bibfnamefont{L.}~\bibnamefont{Wray}},
  \bibinfo{author}{\bibfnamefont{Y.}~\bibnamefont{Xia}},
  \bibinfo{author}{\bibfnamefont{Y.~S.} \bibnamefont{Hor}},
  \bibinfo{author}{\bibfnamefont{R.}~\bibnamefont{Cava}}, \bibnamefont{and}
  \bibinfo{author}{\bibfnamefont{M.~Z.} \bibnamefont{Hasan}},
  \bibinfo{journal}{Nature} \textbf{\bibinfo{volume}{452}},
  \bibinfo{pages}{970} (\bibinfo{year}{2008}).

\bibitem[{\citenamefont{Hsieh et~al.}(2009)\citenamefont{Hsieh, Xia, Qian,
  Wray, Meier, Dil, Osterwalder, Patthey, Fedorov, Lin et~al.}}]{hsieh2}
\bibinfo{author}{\bibfnamefont{D.}~\bibnamefont{Hsieh}},
  \bibinfo{author}{\bibfnamefont{Y.}~\bibnamefont{Xia}},
  \bibinfo{author}{\bibfnamefont{D.}~\bibnamefont{Qian}},
  \bibinfo{author}{\bibfnamefont{L.}~\bibnamefont{Wray}},
  \bibinfo{author}{\bibfnamefont{F.}~\bibnamefont{Meier}},
  \bibinfo{author}{\bibfnamefont{J.~H.} \bibnamefont{Dil}},
  \bibinfo{author}{\bibfnamefont{J.}~\bibnamefont{Osterwalder}},
  \bibinfo{author}{\bibfnamefont{L.}~\bibnamefont{Patthey}},
  \bibinfo{author}{\bibfnamefont{A.~V.} \bibnamefont{Fedorov}},
  \bibinfo{author}{\bibfnamefont{H.}~\bibnamefont{Lin}}, \bibnamefont{et~al.},
  \bibinfo{journal}{Phys. Rev. Lett.} \textbf{\bibinfo{volume}{103}},
  \bibinfo{pages}{146401} (\bibinfo{year}{2009}).

\bibitem[{\citenamefont{H{\"o}ppner et~al.}(2013)\citenamefont{H{\"o}ppner,
  Seiro, Chikina, Fedorov, G{\"u}ttler, Danzenb{\"a}cher, Generalov, Kummer,
  Patil, Molodtsov et~al.}}]{hoppner}
\bibinfo{author}{\bibfnamefont{M.}~\bibnamefont{H{\"o}ppner}},
  \bibinfo{author}{\bibfnamefont{S.}~\bibnamefont{Seiro}},
  \bibinfo{author}{\bibfnamefont{A.}~\bibnamefont{Chikina}},
  \bibinfo{author}{\bibfnamefont{A.}~\bibnamefont{Fedorov}},
  \bibinfo{author}{\bibfnamefont{M.}~\bibnamefont{G{\"u}ttler}},
  \bibinfo{author}{\bibfnamefont{S.}~\bibnamefont{Danzenb{\"a}cher}},
  \bibinfo{author}{\bibfnamefont{A.}~\bibnamefont{Generalov}},
  \bibinfo{author}{\bibfnamefont{K.}~\bibnamefont{Kummer}},
  \bibinfo{author}{\bibfnamefont{S.}~\bibnamefont{Patil}},
  \bibinfo{author}{\bibfnamefont{S.}~\bibnamefont{Molodtsov}},
  \bibnamefont{et~al.}, \bibinfo{journal}{Nature communications}
  \textbf{\bibinfo{volume}{4}}, \bibinfo{pages}{1646} (\bibinfo{year}{2013}).

\bibitem[{\citenamefont{Garner et~al.}(2000)\citenamefont{Garner, Hancock,
  Rodriguez, Schlesinger, Bucher, Fisk, and Sarrao}}]{garner}
\bibinfo{author}{\bibfnamefont{S.~R.} \bibnamefont{Garner}},
  \bibinfo{author}{\bibfnamefont{J.~N.} \bibnamefont{Hancock}},
  \bibinfo{author}{\bibfnamefont{Y.~W.} \bibnamefont{Rodriguez}},
  \bibinfo{author}{\bibfnamefont{Z.}~\bibnamefont{Schlesinger}},
  \bibinfo{author}{\bibfnamefont{B.}~\bibnamefont{Bucher}},
  \bibinfo{author}{\bibfnamefont{Z.}~\bibnamefont{Fisk}}, \bibnamefont{and}
  \bibinfo{author}{\bibfnamefont{J.~L.} \bibnamefont{Sarrao}},
  \bibinfo{journal}{Phys. Rev. B} \textbf{\bibinfo{volume}{62}},
  \bibinfo{pages}{R4778} (\bibinfo{year}{2000}).

\bibitem[{\citenamefont{Dordevic et~al.}(2001)\citenamefont{Dordevic, Basov,
  Dilley, Bauer, and Maple}}]{dordevic}
\bibinfo{author}{\bibfnamefont{S.~V.} \bibnamefont{Dordevic}},
  \bibinfo{author}{\bibfnamefont{D.~N.} \bibnamefont{Basov}},
  \bibinfo{author}{\bibfnamefont{N.~R.} \bibnamefont{Dilley}},
  \bibinfo{author}{\bibfnamefont{E.~D.} \bibnamefont{Bauer}}, \bibnamefont{and}
  \bibinfo{author}{\bibfnamefont{M.~B.} \bibnamefont{Maple}},
  \bibinfo{journal}{Phys. Rev. Lett.} \textbf{\bibinfo{volume}{86}},
  \bibinfo{pages}{684} (\bibinfo{year}{2001}).

\bibitem[{\citenamefont{Mena et~al.}(2005)\citenamefont{Mena, van~der Marel,
  and Sarrao}}]{mena}
\bibinfo{author}{\bibfnamefont{F.~P.} \bibnamefont{Mena}},
  \bibinfo{author}{\bibfnamefont{D.}~\bibnamefont{van~der Marel}},
  \bibnamefont{and} \bibinfo{author}{\bibfnamefont{J.~L.}
  \bibnamefont{Sarrao}}, \bibinfo{journal}{Phys. Rev. B}
  \textbf{\bibinfo{volume}{72}}, \bibinfo{pages}{045119}
  (\bibinfo{year}{2005}).

\bibitem[{\citenamefont{Coleman}(1987)}]{coleman3}
\bibinfo{author}{\bibfnamefont{P.}~\bibnamefont{Coleman}},
  \bibinfo{journal}{Phys. Rev. Lett.} \textbf{\bibinfo{volume}{59}},
  \bibinfo{pages}{1026} (\bibinfo{year}{1987}).

\bibitem[{\citenamefont{Shim et~al.}(2007)\citenamefont{Shim, Haule, and
  Kotliar}}]{shim}
\bibinfo{author}{\bibfnamefont{J.~H.} \bibnamefont{Shim}},
  \bibinfo{author}{\bibfnamefont{K.}~\bibnamefont{Haule}}, \bibnamefont{and}
  \bibinfo{author}{\bibfnamefont{G.}~\bibnamefont{Kotliar}},
  \bibinfo{journal}{Science} \textbf{\bibinfo{volume}{318}},
  \bibinfo{pages}{1615} (\bibinfo{year}{2007}).

\bibitem[{\citenamefont{Weber and Vojta}(2008)}]{weber}
\bibinfo{author}{\bibfnamefont{H.}~\bibnamefont{Weber}} \bibnamefont{and}
  \bibinfo{author}{\bibfnamefont{M.}~\bibnamefont{Vojta}},
  \bibinfo{journal}{Phys. Rev. B} \textbf{\bibinfo{volume}{77}},
  \bibinfo{pages}{125118} (\bibinfo{year}{2008}).

\bibitem[{\citenamefont{Coleman}(1984)}]{coleman2}
\bibinfo{author}{\bibfnamefont{P.}~\bibnamefont{Coleman}},
  \bibinfo{journal}{Phys. Rev. B} \textbf{\bibinfo{volume}{29}},
  \bibinfo{pages}{3035} (\bibinfo{year}{1984}).

\bibitem[{\citenamefont{Travaglini and Wachter}(1984)}]{travaglini}
\bibinfo{author}{\bibfnamefont{G.}~\bibnamefont{Travaglini}} \bibnamefont{and}
  \bibinfo{author}{\bibfnamefont{P.}~\bibnamefont{Wachter}},
  \bibinfo{journal}{Phys. Rev. B} \textbf{\bibinfo{volume}{29}},
  \bibinfo{pages}{893} (\bibinfo{year}{1984}).

\bibitem[{\citenamefont{Kimura et~al.}(1994)\citenamefont{Kimura, Nanba, Kunii,
  and Kasuya}}]{kimura1994}
\bibinfo{author}{\bibfnamefont{S.-i.} \bibnamefont{Kimura}},
  \bibinfo{author}{\bibfnamefont{T.}~\bibnamefont{Nanba}},
  \bibinfo{author}{\bibfnamefont{S.}~\bibnamefont{Kunii}}, \bibnamefont{and}
  \bibinfo{author}{\bibfnamefont{T.}~\bibnamefont{Kasuya}},
  \bibinfo{journal}{Phys. Rev. B} \textbf{\bibinfo{volume}{50}},
  \bibinfo{pages}{1406} (\bibinfo{year}{1994}).

\bibitem[{\citenamefont{Bucher et~al.}(1994)\citenamefont{Bucher, Schlesinger,
  Canfield, and Fisk}}]{bucher}
\bibinfo{author}{\bibfnamefont{B.}~\bibnamefont{Bucher}},
  \bibinfo{author}{\bibfnamefont{Z.}~\bibnamefont{Schlesinger}},
  \bibinfo{author}{\bibfnamefont{P.~C.} \bibnamefont{Canfield}},
  \bibnamefont{and} \bibinfo{author}{\bibfnamefont{Z.}~\bibnamefont{Fisk}},
  \bibinfo{journal}{Phys. Rev. Lett.} \textbf{\bibinfo{volume}{72}},
  \bibinfo{pages}{522} (\bibinfo{year}{1994}).

\bibitem[{\citenamefont{Coleman et~al.}(1996)\citenamefont{Coleman, Schofield,
  and Tsvelik}}]{coleman1}
\bibinfo{author}{\bibfnamefont{P.}~\bibnamefont{Coleman}},
  \bibinfo{author}{\bibfnamefont{A.~J.} \bibnamefont{Schofield}},
  \bibnamefont{and} \bibinfo{author}{\bibfnamefont{A.~M.}
  \bibnamefont{Tsvelik}}, \bibinfo{journal}{Phys. Rev. Lett.}
  \textbf{\bibinfo{volume}{76}}, \bibinfo{pages}{1324} (\bibinfo{year}{1996}).

\bibitem[{\citenamefont{Gell-Mann and Pais}(1955)}]{gellmann}
\bibinfo{author}{\bibfnamefont{M.}~\bibnamefont{Gell-Mann}} \bibnamefont{and}
  \bibinfo{author}{\bibfnamefont{A.}~\bibnamefont{Pais}},
  \bibinfo{journal}{Phys. Rev.} \textbf{\bibinfo{volume}{97}},
  \bibinfo{pages}{1387} (\bibinfo{year}{1955}).

\bibitem[{\citenamefont{Auerbach and Levin}(1986)}]{auerbach}
\bibinfo{author}{\bibfnamefont{A.}~\bibnamefont{Auerbach}} \bibnamefont{and}
  \bibinfo{author}{\bibfnamefont{K.}~\bibnamefont{Levin}},
  \bibinfo{journal}{Phys. Rev. Lett.} \textbf{\bibinfo{volume}{57}},
  \bibinfo{pages}{877} (\bibinfo{year}{1986}).

\bibitem[{\citenamefont{Read et~al.}(1984)\citenamefont{Read, Newns, and
  Doniach}}]{read1}
\bibinfo{author}{\bibfnamefont{N.}~\bibnamefont{Read}},
  \bibinfo{author}{\bibfnamefont{D.~M.} \bibnamefont{Newns}}, \bibnamefont{and}
  \bibinfo{author}{\bibfnamefont{S.}~\bibnamefont{Doniach}},
  \bibinfo{journal}{Phys. Rev. B} \textbf{\bibinfo{volume}{30}},
  \bibinfo{pages}{3841} (\bibinfo{year}{1984}).

\bibitem[{\citenamefont{Newns and Read}(1987)}]{read2}
\bibinfo{author}{\bibfnamefont{D.}~\bibnamefont{Newns}} \bibnamefont{and}
  \bibinfo{author}{\bibfnamefont{N.}~\bibnamefont{Read}},
  \bibinfo{journal}{Advances in Physics} \textbf{\bibinfo{volume}{36}},
  \bibinfo{pages}{799} (\bibinfo{year}{1987}).

\bibitem[{\citenamefont{Semenoff}(1984)}]{semenoff}
\bibinfo{author}{\bibfnamefont{G.~W.} \bibnamefont{Semenoff}},
  \bibinfo{journal}{Phys. Rev. Lett.} \textbf{\bibinfo{volume}{53}},
  \bibinfo{pages}{2449} (\bibinfo{year}{1984}).

\bibitem[{\citenamefont{Saremi and Lee}(2007)}]{saremi}
\bibinfo{author}{\bibfnamefont{S.}~\bibnamefont{Saremi}} \bibnamefont{and}
  \bibinfo{author}{\bibfnamefont{P.~A.} \bibnamefont{Lee}},
  \bibinfo{journal}{Phys. Rev. B} \textbf{\bibinfo{volume}{75}},
  \bibinfo{pages}{165110} (\bibinfo{year}{2007}).

\bibitem[{\citenamefont{Mahan}(2000)}]{mahan}
\bibinfo{author}{\bibfnamefont{G.}~\bibnamefont{Mahan}},
  \emph{\bibinfo{title}{Many Particle Physics}}, Physics of Solids and Liquids
  (\bibinfo{publisher}{Springer}, \bibinfo{year}{2000}), ISBN
  \bibinfo{isbn}{9780306463389}.

\bibitem[{\citenamefont{Paul and Civelli}(2010)}]{paul}
\bibinfo{author}{\bibfnamefont{I.}~\bibnamefont{Paul}} \bibnamefont{and}
  \bibinfo{author}{\bibfnamefont{M.}~\bibnamefont{Civelli}},
  \bibinfo{journal}{Phys. Rev. B} \textbf{\bibinfo{volume}{81}},
  \bibinfo{pages}{161102} (\bibinfo{year}{2010}).

\bibitem[{\citenamefont{Ye et~al.}(2012)\citenamefont{Ye, Eremeev, Kuroda,
  Krasovskii, Chulkov, Takeda, Saitoh, Okamoto, Zhu, Miyamoto
  et~al.}}]{kimura2012}
\bibinfo{author}{\bibfnamefont{M.}~\bibnamefont{Ye}},
  \bibinfo{author}{\bibfnamefont{S.}~\bibnamefont{Eremeev}},
  \bibinfo{author}{\bibfnamefont{K.}~\bibnamefont{Kuroda}},
  \bibinfo{author}{\bibfnamefont{E.}~\bibnamefont{Krasovskii}},
  \bibinfo{author}{\bibfnamefont{E.}~\bibnamefont{Chulkov}},
  \bibinfo{author}{\bibfnamefont{Y.}~\bibnamefont{Takeda}},
  \bibinfo{author}{\bibfnamefont{Y.}~\bibnamefont{Saitoh}},
  \bibinfo{author}{\bibfnamefont{K.}~\bibnamefont{Okamoto}},
  \bibinfo{author}{\bibfnamefont{S.}~\bibnamefont{Zhu}},
  \bibinfo{author}{\bibfnamefont{K.}~\bibnamefont{Miyamoto}},
  \bibnamefont{et~al.}, \bibinfo{journal}{Physical Review B}
  \textbf{\bibinfo{volume}{85}}, \bibinfo{pages}{205317}
  (\bibinfo{year}{2012}).

\end{thebibliography}

\end{document}